\documentclass[a4paper,pra,twocolumn,superscriptaddress]{revtex4-2}

\usepackage{amssymb}
\usepackage{amsmath}
\usepackage{epsfig}
\usepackage{color}
\usepackage{graphics, graphicx}
\usepackage{bbold}
\usepackage{psfrag}
\usepackage{mathcomp}
\usepackage{subfigure}
\usepackage{verbatim}
\usepackage{blindtext}
\usepackage{float}
\usepackage[utf8]{inputenc}
\usepackage[T1]{fontenc}
\usepackage{supertabular}
\usepackage{longtable}
\usepackage[colorlinks,citecolor=blue]{hyperref}

\begin{document}

\date{\today}
\title{Spontaneous breaking of continuous scale invariance and Efimovian-like dynamics in a driven-dissipative system}
\author{Zehui Wu}
\affiliation{College of Physics, Sichuan University, Chengdu 610065, China}
\author{Jian Yu}
\email{yujian@sztu.edu.cn}
\affiliation{Center for Intense Laser Application Technology, College of Engineering Physics, and Guangdong Engineering Technology Research Center for High-Power Ultrafast Laser and X-Ray Equipment, Shenzhen Technology University, Shenzhen 518118, China}
\author{Jian-Song Pan}
\email{panjsong@scu.edu.cn}
\affiliation{College of Physics, Sichuan University, Chengdu 610065, China}
\affiliation{Key Laboratory of High Energy Density Physics and Technology of Ministry of Education, Sichuan University, Chengdu 610065, China}

\begin{abstract}
The Efimov effect manifests discrete scale invariance through a geometric series of three-body bound states. Analogous discrete scale symmetry has been observed in the Efimovian expansion of scale-invariant strongly interacting Fermi gases. However, it is unclear whether strong correlation is necessary. Here we investigate the spontaneous breaking of continuous scale invariance in a driven-dissipative two-level system whose driving and dissipation parameters scale as
\(1/t\). By transforming to logarithmic time, we show that the evolution is governed by a non-Hermitian superoperator whose spectral decomposition reveals the emergence of a pair of eigenmodes with opposite real parts and equal imaginary parts. This signals the spontaneous breaking of continuous scale invariance and gives rise to Efimovian-like oscillations: log-periodic decay oscillations obeying discrete scaling laws. The onset of oscillations coincides with a PT-like-symmetry
breaking transition in logarithmic time. Our work implies that the spontaneous breaking of continuous temporal scale symmetry and the emergence of Efimovian-like dynamics can be observed in a minimal driven-dissipative system without
requiring strong correlations.
\end{abstract}

\maketitle

\section{Introduction}
\label{sec:intro}
The interplay between scale invariance and quantum dynamics has long been a fascinating theme in modern physics, ranging from critical phenomena~\cite{wilson1971renormalization, hohenberg1977theory, wegner2005critical, fisher1998renormalization, yang2007quantum, leonel2025scaling} and few-body physics~\cite{efimov1970yad, efimov1970physics, efimov1972level, efimov1973energy, bedaque1999renormalization, efimovphysicsreview}, to unitary quantum gases~\cite{o2002observation,
ho2004universal, castin2004exact, kraemer2006evidence, braaten2006universal, knoop2009observation, zaccanti2009observation, pollack2009universality, barontini2009observation, chin2010feshbach, ferlaino2011efimov, pires2014observation, deng2015observation, maier2015efimov, deng2017observation, sun2017visualizing, ulmanis2016heteronuclear, ulmanis2016nsr, xie2020observation, wang2024scale, lippi2024experimental}. A landmark discovery in this context was made by Efimov in
1970~\cite{efimov1970yad, efimov1970physics, efimov1972level, efimov1973energy}, who revealed that three identical bosons interacting via short-range resonant interactions can form an infinite series of bound states whose energies and sizes obey a geometric progression. This remarkable discrete scale invariance, characterized by a universal scaling factor \(e^{2\pi/|s_0|}\approx 515\) for identical bosons, emerges from an effective long-range three-body attraction that is
itself a consequence of the resonant two-body interactions. The Efimov effect, as it came to be known, represents a paradigmatic example of how discrete scale symmetry can arise from continuous scale invariance when an appropriate boundary condition is imposed. Over the past five decades, Efimov physics has been extensively investigated both theoretically and experimentally in cold atoms~\cite{kraemer2006evidence, braaten2006universal,   knoop2009observation,
zaccanti2009observation, pollack2009universality,barontini2009observation, ferlaino2011efimov, pires2014observation, maier2015efimov, ulmanis2016heteronuclear, ulmanis2016nsr, sun2017visualizing, xie2020observation,lippi2024experimental}.

Remarkably, the concept of discrete scale invariance is not confined to static few-body bound states but also manifests itself in far-from-equilibrium dynamics. In a seminal series of experiments, Deng \emph{et. al.}~\cite{deng2015observation,deng2017observation} demonstrated that the expansion dynamics of a scale-invariant ultracold Fermi gas of \(^6\)Li atoms in a harmonically trapped potential exhibits striking plateaus when the trap frequency is ramped down as
\(\omega(t)=1/(\sqrt{\lambda}t)\). Specifically, the mean-square cloud size \(\langle R^2(t)\rangle\) follows a log-periodic function of time, with the expansion temporarily stalling at discrete instants \(t_n = t_0 e^{2\pi n/s_0}\). This ``Efimovian expansion'' represents a temporal analog of the Efimov effect: the continuous scaling symmetry of the underlying hydrodynamic equations is spontaneously broken down to a discrete scaling symmetry by the initial conditions
imposed at \(t=t_0\). The universal scaling factor and the log-periodic oscillations bear a striking mathematical resemblance to those of the original Efimov trimers, thereby establishing a deep connection between few-body bound states and nonequilibrium dynamics of many-body systems. Subsequent work \cite{deng2017observation} extended this phenomenon to the ``super-Efimovian'' regime, where the trap frequency follows a more complex time dependence involving logarithmic
factors, leading to double-log-periodic oscillations that mirror the super-Efimov effect predicted for fermions in two dimensions.

Parallel to these developments, the concept of time crystals, where the temporal translation symmetry spontaneously breaks, has attracted intense interest in recent years~\cite{shapere2012classical, wilczek2012quantum, sacha2018time}. While continuous time crystals in closed systems with short-range interaction are thought to be forbidden \citep{bruno2013impossibility,watanabe2015absence}, discrete time crystals in periodically driven many-body strongly-correlated systems
\citep{zhang2017observation,choi2017observation} and continuous time crystals in driven-dissipative systems \citep{keßler2021observation} have been demonstrated. The profound interest in the spontaneous breaking of temporal translational symmetry naturally leads us to ask: what is the underlying mechanism for the spontaneous breaking of continuous scale symmetry in time?

Inspired by the driven-dissipative time crystal~\cite{sacha2018time}, we investigate a simple driven-dissipative two-level atom whose driving and dissipation parameters scale with time as \(1/t\). This model possesses no spatial degrees of freedom and has only a temporal degree of freedom, which is sufficient for continuous temporal scale invariance. We find that this system exhibits the spontaneous breaking of continuous scale symmetry, giving rise to Efimovian-like
oscillations: the normalized excited-state population displays log-periodic decay oscillations in logarithmic time, with the oscillation period satisfying the same discrete scale invariance as in the Efimovian expansion. By mapping the master equation to an autonomous form via the logarithmic time transformation \(\tau = \ln(t/t_0)\), we show that the dynamics is governed by a time-independent non-Hermitian superoperator. The initial condition at \(t=t_0\) plays the role of a
``three-body parameter'' in the Efimov problem, breaking the continuous scaling symmetry down to a discrete subgroup. This mechanism is conceptually analogous to the origin of the Efimov effect in three-body systems, where the continuous scale invariance of the zero-range theory is broken by a three-body boundary condition.

Furthermore, we uncover a connection between Efimovian-like oscillations and the spontaneous breaking of parity-time (PT) symmetry~\cite{bender1998real}. In the undriven limit, the system resides in the PT-broken phase, where all eigenvalues of the superoperator are imaginary, leading to monotonic decay without oscillations. As the driving strength is increased beyond a critical threshold, the system undergoes a PT-symmetry breaking transition into the PT-symmetric phase,
where a pair of eigenvalues acquires opposite real parts and controls the oscillatory dynamics. This transition precisely coincides with the appearance of log-periodic oscillations, indicating that Efimovian-like oscillations are a dynamical manifestation of PT-symmetry breaking in the logarithmic time domain. To examine the role of interatomic coupling, we extend the model to one-dimensional chains with nearest-neighbor density-density interactions. Comparing chains of different lengths, we find that the normalized average excitation is essentially independent of the number of atoms. By analyzing the dynamics under white noise, we find that the oscillation is robust against small perturbations.

The remainder of this paper is organized as follows. In Sec.~\ref{sec:model}, we introduce the two-level model and perform the logarithmic time transformation that renders the master equation autonomous. In Sec.~\ref{sec:biorthogonal}, we present the superoperator formalism, the two-dimensional density-matrix expansion, and the biorthogonal projection method for different initial atomic states. We also discuss the eigenvalue spectrum and construct the dynamical phase
diagram. In Sec.~\ref{sec:interaction}, we extend the model to finite chains with inter-atom interaction, show that the normalized dynamics is essentially independent of the number of atoms, and analyze the effect of white noise. Finally, Sec.~\ref{sec:conclusion} presents our conclusions and additional discussion.

\section{Model and phase diagram}
\label{sec:model}

We consider a driven-dissipative two-level atom with parameters scaling as $1/t$. In the interaction picture, the density matrix $\rho(t)$ evolves as
\begin{equation}\label{eq:Lindblad master eq}
i\frac{d\rho}{dt}=\left[H(t),\,\rho\right]+i\kappa(t)\left(2\sigma^-\rho\sigma^+-\sigma^+\sigma^-\rho-\rho\sigma^+\sigma^-\right)/2,
\end{equation}
with the driving Hamiltonian
\begin{equation}\label{eq:Hamiltonian}
    H(t)=i\eta(t)(\sigma^+-\sigma^-).
\end{equation}
Here \(\sigma^+=|e\rangle\langle g|\) and \(\sigma^-=|g\rangle\langle e|\) are the raising and lowering operators of the two-level atom, \(|e\rangle\) and \(|g\rangle\) denote the excited and ground states, \(\eta(t)\) is the driving amplitude, and \(\kappa(t)\) is the dissipation rate. The excited-state population is \(p_e(\tau)=\mathrm{Tr}[\rho(\tau)\sigma^+\sigma^-]\). In the figures for the single-atom problem we plot the signal
\begin{equation}\label{eq:normalized population}
    n(\tau)\equiv \,\mathrm{Tr}[\rho(\tau)\sigma^+\sigma^-]=p_e(\tau),
\end{equation}
so that the ground state, the excited state, and the maximally mixed state correspond to \(n=0\), \(1\), and \(1/2\), respectively. We focus on the situation with continuous scale symmetry, $t\rightarrow \lambda t, \rho\rightarrow \lambda \rho$, which is satisfied when the parameters scale inversely with time,
\begin{equation}\label{parameters}
    \eta(t)=\frac{\eta_0}{t},\quad \kappa(t)=\frac{\kappa_0}{t}.
\end{equation}
For a two-level atom, the time-dependent drive may be realized by controlling the amplitude of a resonant laser. The atomic operators satisfy \(\{\sigma^+,\sigma^-\}=1\) and \((\sigma^+)^2=(\sigma^-)^2=0\).

\begin{figure}
    \centering
    \includegraphics[width=0.8\linewidth]{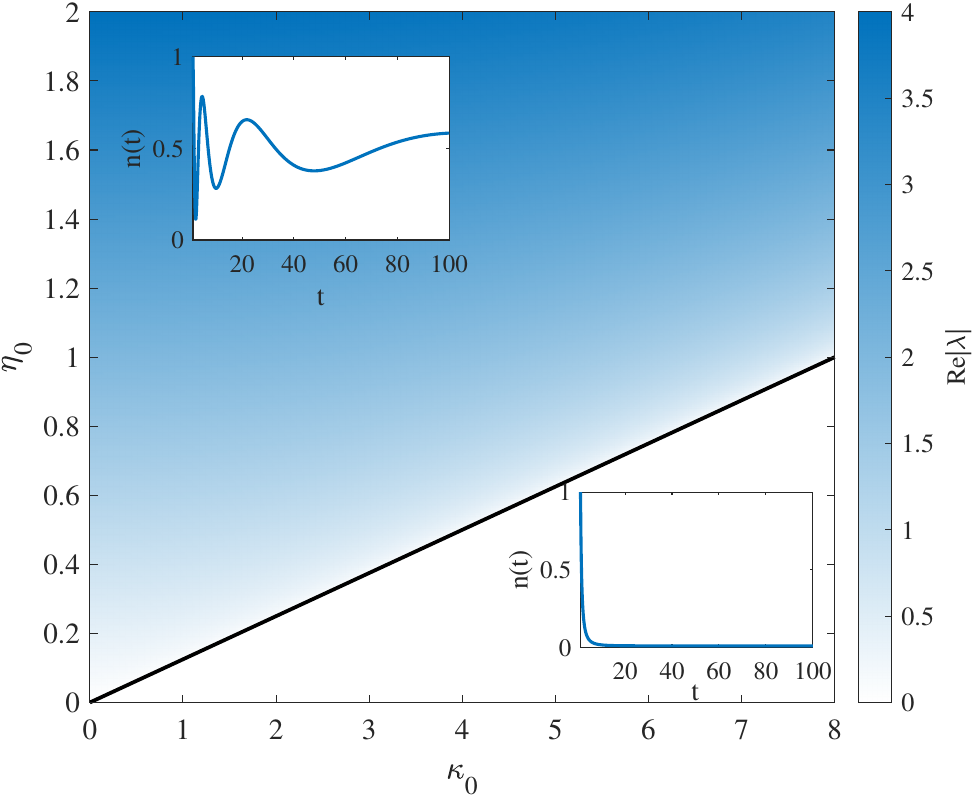}
    \caption{The dynamical phase diagram of the driven-dissipative two-level atom in the $\eta_0$-$\kappa_0$ parameter plane. The color scale shows the eigenvalue pair \(|\mathrm{Re}\,\lambda_\pm|\) that dominate the oscillation, and the black curve separates the oscillatory and non-oscillatory regions. The upper-left inset shows the normalized population \(n(t)=p_e(t)\) at a representative oscillatory point, displaying damped log-periodic oscillations. The lower-right inset shows the
non-oscillatory evolution at a representative point on the other side of the boundary, where \(n(t)\) decays monotonically toward its steady value.}
    \label{fig:phase}
\end{figure}

We next introduce the logarithmic time variable $\tau=\ln(t/t_0)$, where $t_0$ is an arbitrary initial time. Through this transformation, the generator becomes time-independent,
\begin{equation}\label{autonomous form}
    i\frac{d\rho}{d\tau}=\left[H_0,\,\rho\right]+i\kappa_0\left(2\sigma^-\rho\sigma^+-\sigma^+\sigma^-\rho-\rho\sigma^+\sigma^-\right)/2,
\end{equation}
with \(H_0=i\eta_0(\sigma^+-\sigma^-)\). Equation~(\ref{autonomous form}) is the equation of motion for a driven-dissipative two-level atom in logarithmic time~\cite{rivas2012open,ficek2014quantum}.

Expanding the density matrix in the atomic basis \(\{\rho_{ee},\rho_{eg},\rho_{ge},\rho_{gg}\}\), we obtain
\begin{equation}\label{density matrix elements}
\begin{split}
    \frac{d\rho_{ee}}{d\tau}&=-\kappa_0\rho_{ee}+\eta_0(\rho_{eg}+\rho_{ge}),\\
    \frac{d\rho_{eg}}{d\tau}&=-\frac{\kappa_0}{2}\rho_{eg}+\eta_0(\rho_{gg}-\rho_{ee}),\\
    \frac{d\rho_{ge}}{d\tau}&=-\frac{\kappa_0}{2}\rho_{ge}+\eta_0(\rho_{gg}-\rho_{ee}),\\
    \frac{d\rho_{gg}}{d\tau}&=\kappa_0\rho_{ee}-\eta_0(\rho_{eg}+\rho_{ge}).
\end{split}
\end{equation}
This closed set of equations is the foundation for the numerical study below. All coefficients are real, and the reduced density matrix is fully determined by these four real equations with \(\mathrm{Tr}\,\rho=1\).

To show the dynamics, we numerically solve Eq.~(\ref{density matrix elements}). As shown in Fig.~\ref{fig:phase}, when the scaled driving strength \(\eta_0\) is below (above) a threshold proportional to the decay rate \(\kappa_0\), the normalized population \(n(\tau)=p_e(\tau)\) exhibits monotonic (oscillatory) decay. In logarithmic time coordinates, \(n(\tau)\) shows periodic decay oscillations in the oscillatory phase. The oscillation in the original time becomes an
Efimovian-like oscillation with geometrically increasing periods, as shown in the insets of Fig.~\ref{fig:phase}. This log-periodicity is the signature of the spontaneous breaking of continuous scale symmetry into a discrete scale symmetry.

\begin{figure}
    \centering
    \includegraphics[width=1.0\linewidth]{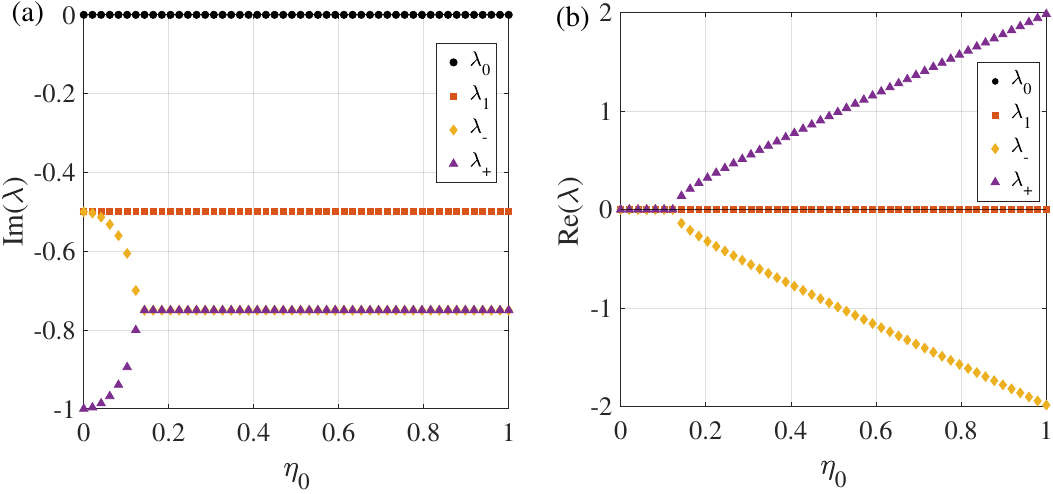}
    \caption{The spectra of $\mathcal{H}$ as functions of $\eta_0$ at fixed $\kappa_0=1$. The imaginary parts are shown in (a) and the real parts in (b). The eigenvalues $\lambda_\pm$ are purely imaginary below the critical drive, become degenerate at $\eta_0\simeq\kappa_0/8$, and acquire opposite real parts above it. The real parts are nondegenerate in the large-\(\eta_0\) regime (right). This transition implies a PT phase transition of $i\mathcal{H}$.}
    \label{fig:spectra}
\end{figure}

\section{Spectral analysis}
\label{sec:biorthogonal}

To obtain the spectra characterizing the dynamics, we vectorize the density matrix by stacking its columns into a column vector $\vec\rho$ of length \(D^2\), with \(D=2\) for the atomic Hilbert space~\cite{yi2001effective,navarrete2015open}. The master equation takes the form $i\,d\vec\rho/d\tau=\mathcal{H}\vec\rho$, where
\begin{equation}\label{eq:superoperator}
\begin{split}
\mathcal{H}={}&I\otimes H_0-H_0^T\otimes I\\
&+\frac{i\kappa_0}{2}\left(2\sigma^-\otimes\sigma^--\sigma^+\sigma^-\otimes I-I\otimes\sigma^+\sigma^-\right),
\end{split}
\end{equation}
and \(I\) is the \(2\times2\) identity matrix. Because \(D=2\), \(\mathcal{H}\) is a \(4\times4\) matrix.

To understand the origin of the Efimovian-like oscillations, we turn to the spectral decomposition of $\mathcal{H} $. Because the matrix $\mathcal{H}$ is non-Hermitian, its right eigenvectors $v_j$ (satisfying $\mathcal{H}v_j=\lambda_jv_j$, $\lambda_j$ is the j-th eigenvalue of $\mathcal{H}$) are not mutually orthogonal. In order to expand any eigenstates, we introduce the left eigenvectors $u_j$. $u_j$ is defined by $u_j^\dagger\mathcal{H}=\lambda_ju_j^\dagger$. The left
eigenvector and the right eigenvector are normalized to satisfy the biorthonormality condition $u_j^\dagger v_l=\delta_{jl}$.

\begin{figure}
    \centering
    \includegraphics[width=1.0\linewidth]{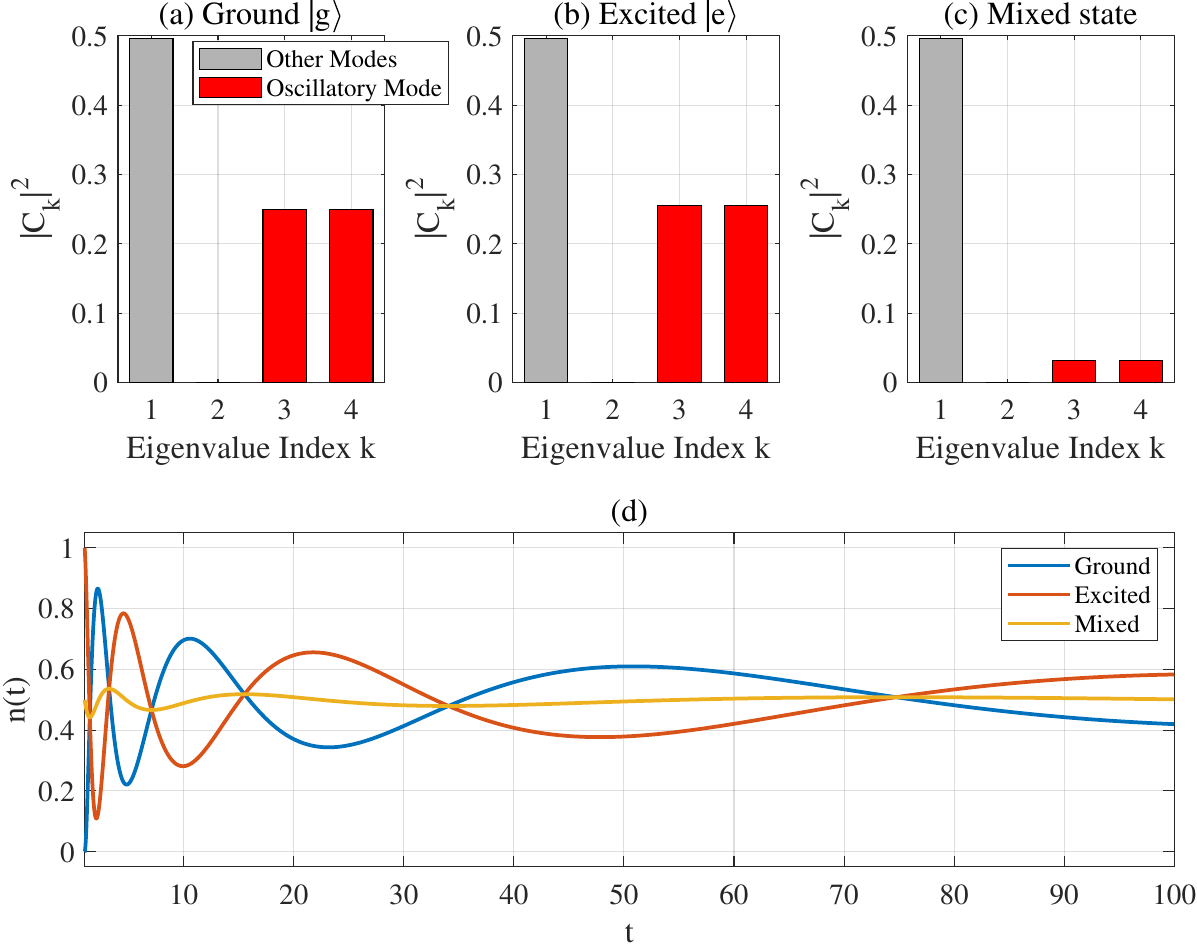}
    \caption{Projection weights on the biorthogonal eigenmodes for three initial atomic states: the ground state \(|g\rangle\langle g|\) (a), the excited state \(|e\rangle\langle e|\) (b), and the maximally mixed state \((|e\rangle\langle e|+|g\rangle\langle g|)/2\) (c). The horizontal axis labels the eigenvalue index, and the dominant oscillatory modes are marked in red. Panel (d) shows the corresponding excited-state population \(n(t)=p_e(t)\) for all three initial states.}
    \label{fig:projection}
\end{figure}

Any density matrix vector can be expanded as
\begin{equation}\label{expansion desity vector}
    \vec \rho(\tau)=\sum_j(c_je^{-i \lambda_j \tau})v_j,
\end{equation}
where the expansion coefficients $c_j$ are obtained from the left eigenvectors and the initial density matrix, $c_j=u_j^\dagger\vec\rho(0)$. The projection weights $|c_j|^2$ measure how strongly the \(j\)-th eigenmode is initially populated.

For the Hamiltonian studied in this article, apart from the steady eigenvalue with zero imaginary part, the products of $i$ and eigenvalues form conjugate pairs $i\lambda_k^{\pm}=\pm ia_k- b_k$ with $b_k<0$, i.e., the eigenvalues are symmetric about the imaginary axis (see Fig.~\ref{fig:spectra}). We find \(\lambda_0=0\) and \(\lambda_1=-i\kappa_0/2\). When $\eta_0$ is below (above) the critical value, the remaining pair has $a_k=0$ ($a_k\neq 0$), and the system shows monotonic (oscillatory) damping behavior.

The eigenvalues of $\mathcal{H}$ can be ordered by their imaginary parts. When \(\eta_0\) is smaller than the critical value, the real parts of the excited eigenvalues are exactly zero. When \(\eta_0\) exceeds the critical value, a pair of real parts with opposite signs emerges, marking the onset of the log-periodic oscillation. The transition from a purely imaginary spectrum to a pair of eigenvalues with real parts of opposite sign is reminiscent of the breaking of the PT
symmetry of \(i\mathcal{H}\), which corresponds to the emergence of eigenvalues symmetric about the imaginary axis~\cite{bender1998real}.

The largest imaginary part is always zero and corresponds to the steady state. The non-oscillatory mode \(\lambda_1=-i\kappa_0/2\) decays more slowly than the oscillatory pair, but its projection on the symmetric initial states considered below is small compared with that of the oscillatory pair. The observed oscillation is therefore controlled by \(\lambda_\pm=\pm a+i b\) with \(a\neq0\). It produces log-periodic oscillations with period \(T=2\pi/a\) in logarithmic time.
The normalized population is approximately
\begin{equation}\label{approximate equation}
    n(\tau)\approx n_0+Ae^{-|b|\tau}\cos(a\tau+\varphi),
\end{equation}
where \(n_0\) is the steady-state value, \(A\) is the oscillation amplitude, and \(\varphi\) is the phase. In the non-oscillatory region all eigenvalues are purely imaginary and the signal decays monotonically.

Because the expansion coefficients $c_k$ differ, different initial atomic states can exhibit different oscillation visibility. Fig.~\ref{fig:projection} shows the projection weights together with the corresponding time evolution for the ground, excited, and maximally mixed states. Although the initial projections are different, the long-time oscillations are governed by the same conjugate pair of eigenmodes, as shown in Fig.~\ref{fig:projection}(d).

\section{Interactions and noises}
\label{sec:interaction}
Having established the single-atom mechanism, we extend the model to a one-dimensional chain of \(N\) identical two-level atoms at the sites \(j=1,\ldots,N\). Adjacent sites are coupled by density-density interactions. With \(n_j=\sigma_j^+\sigma_j^-=|e_j\rangle\langle e_j|\), the Hamiltonian in the interaction picture is
\begin{equation}\label{eq:chain Hamiltonian}
    H(t)=i\eta(t)\sum_{j=1}^{N}\left(\sigma_j^+-\sigma_j^-\right)
    +\frac{V_0}{t}\sum_{j=1}^{N-1}n_jn_{j+1},
\end{equation}
where \(V_0\) is a constant interaction coefficient. Its \(V_0/t\) scaling preserves the continuous temporal scale symmetry and keeps the logarithmic-time description autonomous.

\begin{figure}[t!]
    \centering
    \includegraphics[width=1.0\linewidth]{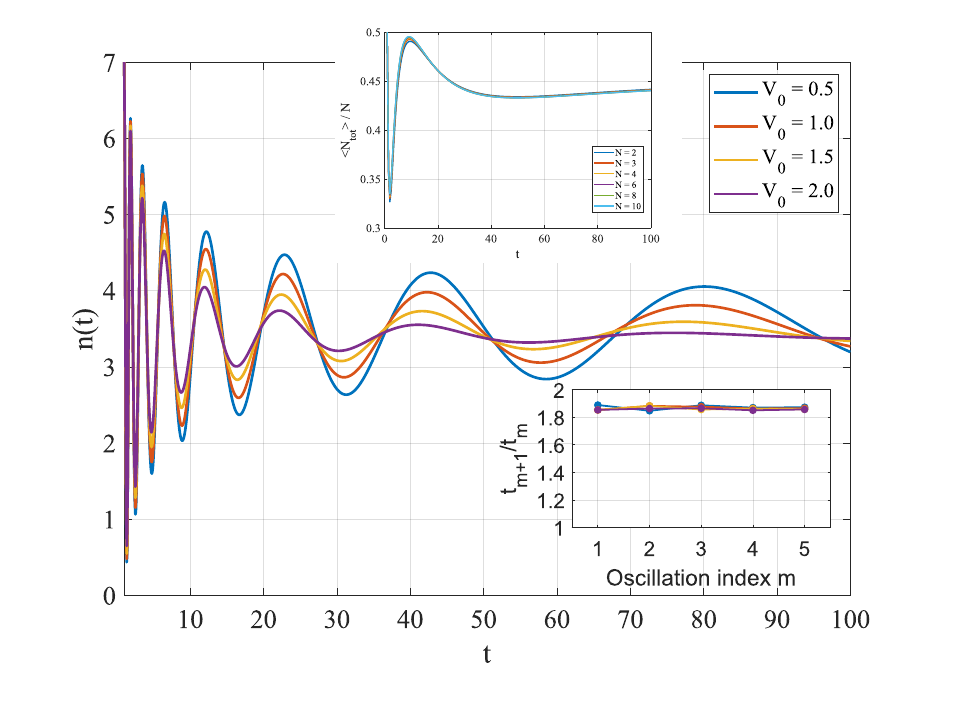}
    \caption{Interaction and atom-number dependence of the log-periodic dynamics. The main panel shows \(n(t)\) for the representative three-atom chain with interaction strengths \(V_0=0.5\), \(1.0\), \(1.5\), and \(2.0\). The upper inset compares the average excitation \(\langle N_{\rm tot}\rangle/N\) for different numbers of atoms, \(N=2,3,4,6,8,10\); after division by \(N\), the curves nearly collapse onto one another. The lower inset shows the ratio \(t_{m+1}/t_m\) of adjacent oscillation periods for the representative chain. The ratio is nearly constant across
different interaction strengths and oscillation indices, confirming that the oscillation times form a geometric sequence.}
    \label{fig:scaled_interaction}
\end{figure}

Fig.~\ref{fig:scaled_interaction} first tests the dependence on the number of atoms. In the upper inset, the average excitation curves for \(N=2,3,4,6,8,10\) collapse after division by \(N\) over the range shown. The dynamics is therefore controlled by the excitation density rather than by the absolute number of atoms, and the choice \(N=3\) in the main panel and in the spectral analysis is merely representative. For all interaction strengths shown, the population exhibits sustained log-periodic oscillations with nearly the same period in logarithmic time, while the interaction changes the oscillation amplitude and phase. The lower inset confirms that the ratio between neighboring oscillation periods is almost independent of both \(V_0\) and the oscillation index, so the oscillation times obey a
geometric progression. Thus the interaction affects the detailed waveform but does not alter the discrete scaling structure.

\begin{figure}[t!]
    \centering
    \includegraphics[width=1.0\linewidth]{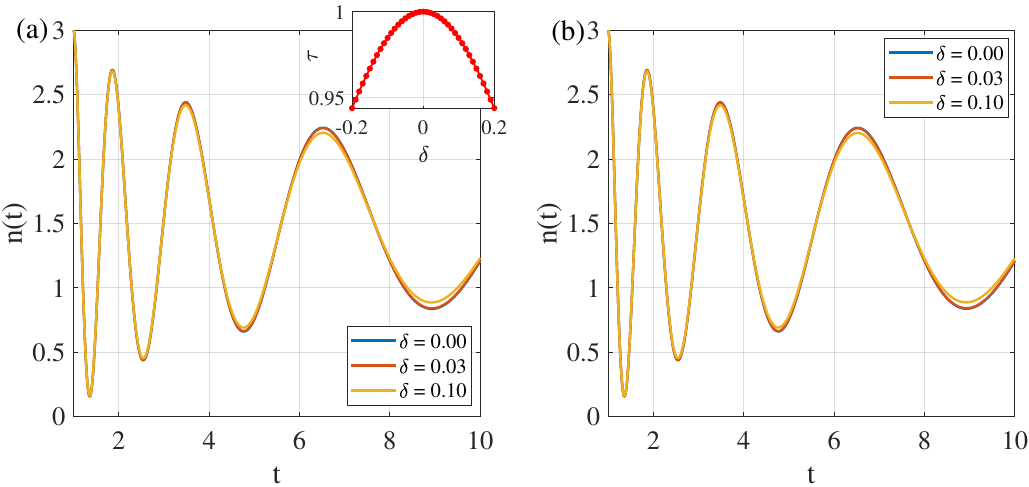}
    \caption{Effect of Gaussian white noise on the Efimovian-like oscillation. (a) The noise is scaled with time, \(\eta_{\rm s}(t)=\eta_0/t+\delta\xi(t)/t\). (b) The same white noise is added without this scaling, \(\eta_{\rm ns}(t)=\eta_0/t+\delta\xi(t)\). Curves correspond to fluctuation strengths \(\delta=0.00\), \(0.03\), and \(0.10\). The inset of (a) shows \(r(\delta)=1-\int_0^{20}dt\,|n(t,\delta)-n(t,0)|/\int_0^{20}dt\,n(t,0)\), which measures the deviation
of \(n(t)\) from the unperturbed evolution. For small \(\delta\), the curves in both protocols almost collapse onto the unperturbed result; even for stronger noise, the oscillation remains qualitatively unchanged. This indicates that the Efimovian-like oscillation is insensitive to white noise.}
    \label{fig:noise}
\end{figure}

To test whether this stability depends on the \(1/t\) scaling, we add zero-mean Gaussian white noise \(\xi(t)\) to the drive in two ways. In protocol (a), the noise is scaled with time as $\eta_{\rm s}(t)=\frac{\eta_0}{t}+\frac{\delta}{t}\xi(t)$,
whereas in protocol (b) the same white-noise term is added directly in the original time, $\eta_{\rm ns}(t)=\frac{\eta_0}{t}+\delta\xi(t)$.
Here \(\delta\) is the fluctuation strength. Fig.~\ref{fig:noise} compares the two protocols for \(\delta=0.00\), \(0.03\), and \(0.10\). For small \(\delta\), the noisy curves are nearly indistinguishable from the unperturbed curve in both cases. The inset of Fig.~\ref{fig:noise}(a) further shows that \(r(\delta)\) varies only weakly over
\(-0.2\le\delta\le 0.2\). The two protocols exhibit no qualitative difference in the oscillation signal, indicating that preserving the \(1/t\) scaling of the noise is not required for the oscillation to survive. These results show that the Efimovian-like oscillation is insensitive to white noise on the timescales considered.

\section{Conclusion and discussion}
\label{sec:conclusion}
We have shown that a driven-dissipative two-level system with parameters scaling as $1/t$ exhibits the spontaneous breaking of continuous temporal scale invariance and the emergence of Efimovian-like oscillations. The continuous scale
symmetry is broken down to a discrete subgroup by the initial condition at \(t=t_0\), which plays the role of the three-body parameter in the Efimov problem, and the resulting log-periodic dynamics obeys the same discrete scaling law as the Efimov effect. We identify the onset of these oscillations with a PT-symmetry breaking transition in logarithmic time: the oscillations emerge when a pair of purely imaginary eigenvalues of the non-Hermitian superoperator acquires
opposite real parts. We have also examined finite chains with nearest-neighbor interactions and found that the normalized dynamics is essentially independent of the number of atoms; including these interactions does not change the central results. The noise tests further show that the log-periodic structure survives both scaled and non-scaled perturbations. Our results demonstrate that the spontaneous breaking of continuous temporal scale symmetry and Efimovian-like dynamics can arise in a minimal driven-dissipative system without requiring strong correlations or resonant two-body interactions.

The dynamics studied in this article have similarities with the Efimovian expansion observed in an ultracold Fermi gas~\cite{deng2015observation}. In that experiment, the mean-square cloud size obeys a third-order differential equation and its solution can be expressed as
$\langle R^2(t)\rangle\approx t[1-A\cos(s_0\ln t+\phi)]$.
The expansion stalls at times $t_n=t_0e^{2\pi n/s_0}$, creating observable plateaus where the $\frac{d\langle R^2\rangle}{dt}=0$. In our system, the normalized population follows an analogous form to $\langle R^2(t)\rangle$. The extrema are separated by a constant interval $\Delta\tau=\pi/a$ in logarithmic time. The plateaus in the Efimovian expansion correspond to the peaks and valleys of the population signal.

\section*{Acknowledgments}
We thank Silvio Flocco for helpful discussions. This work was supported by the National Natural Science Foundation of China (NSFC) under Grant No. 12574297, the Natural Science Foundation of Sichuan Province under Grant No. 2025ZNSFSC0058, the Fundamental Research Funds for the Central Universities under Grant No. YJ202212, the National Key R$\&$D Program of China under Grants No. 2024YFF0508503, the National Natural Science Foundation of China (NSFC) under Grant No. 12405247, and the Guangdong Provincial Department of
Young Innovative Talents  Project under Grant No. 2024KQNCX038.

\end{document}